\documentclass[12pt]{article}
\usepackage{epsf}
\usepackage[utf8]{inputenc}
\usepackage[english]{babel}

\begin{document}

\title{{\bf The hard pomeron intercept and the data on the proton unpolarized structure function}}
\author{A.A. Godizov\thanks{E-mail: anton.godizov@gmail.com}\\
{\small {\it Institute for High Energy Physics, 142281 Protvino, Russia}}}
\date{}
\maketitle

\begin{abstract}
It is demonstrated that the hard pomeron intercept value can be estimated directly from the data on the proton unpolarized structure function: 
$\alpha_{\rm P}^h(0)-1=0.317\pm 0.028$.
\end{abstract}

\section*{Introduction}

The aim of this note is to demonstrate that such an important quantity as the hard pomeron intercept (which governs the behavior of hadronic cross-sections at ultra-high 
energies) can be extracted from the experimental data directly and, as well, to provide an explicit estimation of the corresponding value. 

\begin{figure}[ht]
\epsfxsize=8cm\epsfysize=8cm\epsffile{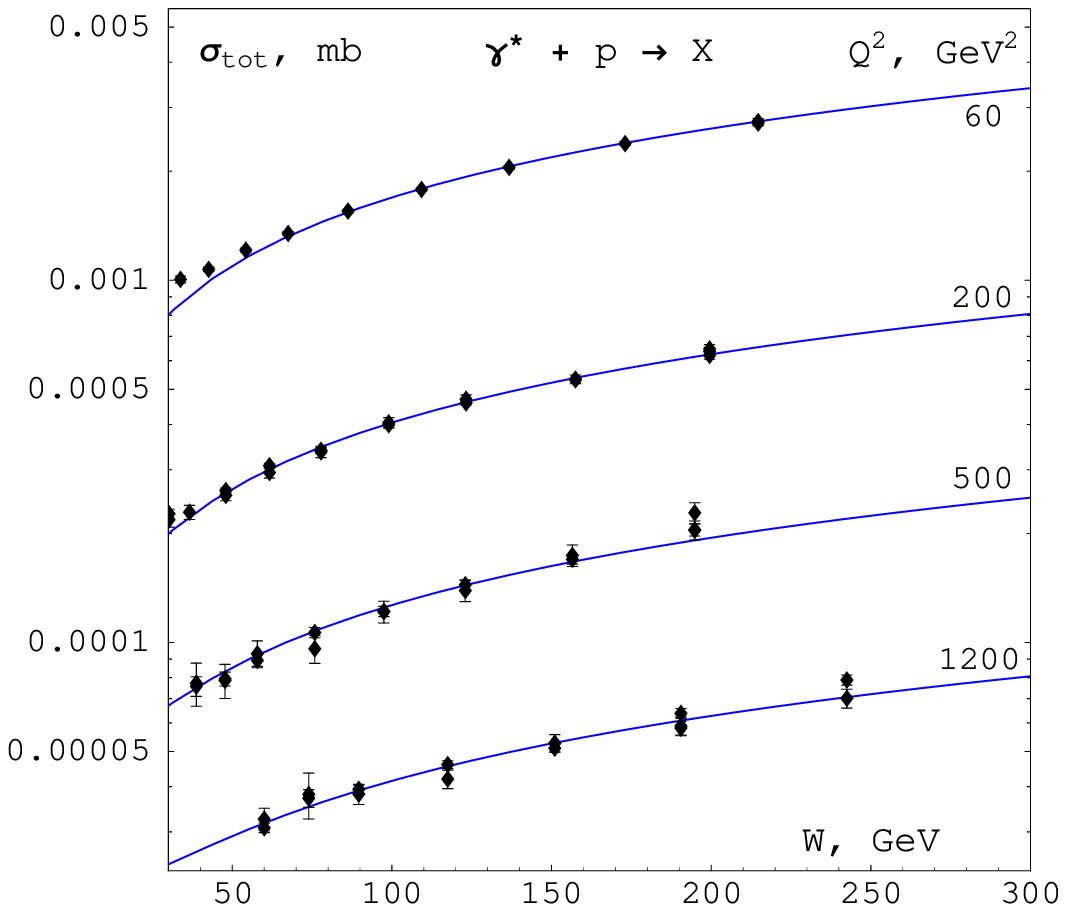}
\vskip -8cm
\hskip 8.5cm
\epsfxsize=8cm\epsfysize=8cm\epsffile{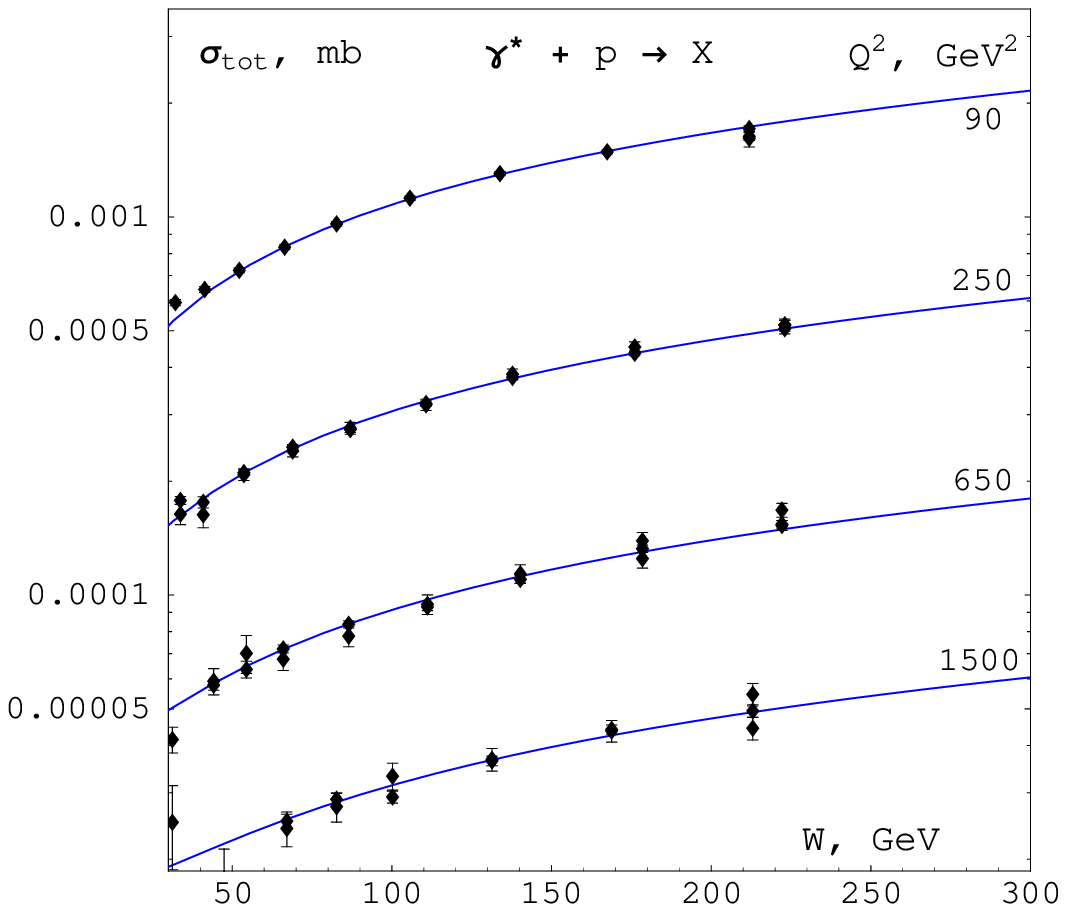}
\vskip -0.6cm
\epsfxsize=8cm\epsfysize=8cm\epsffile{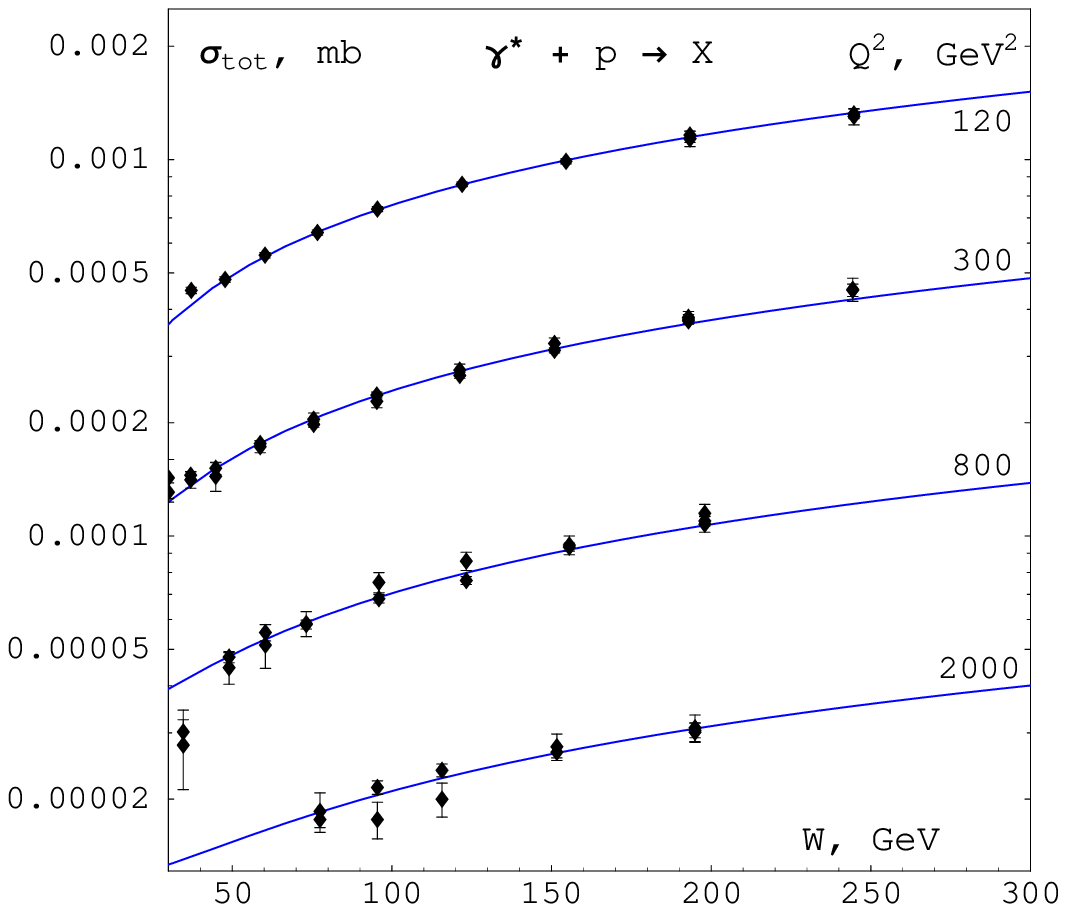}
\vskip -8cm
\hskip 8.5cm
\epsfxsize=8cm\epsfysize=8cm\epsffile{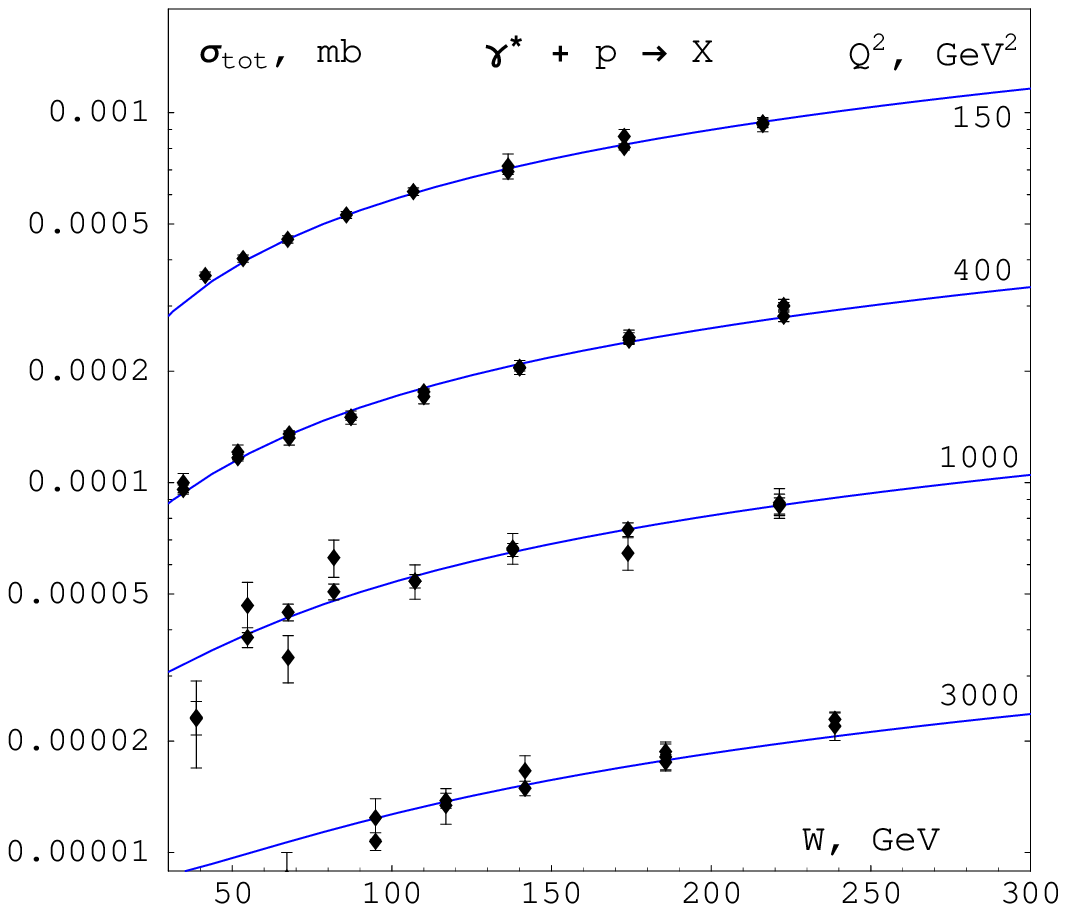}
\vskip -0.5cm
\caption{The description of the $\gamma^*p$ total cross-sections at high values of $Q^2$ by means of expression (\ref{power}) with $\delta=0.317$.}
\label{risstr}
\end{figure}

\begin{figure}[ht]
\begin{center}
\epsfxsize=8cm\epsfysize=8cm\epsffile{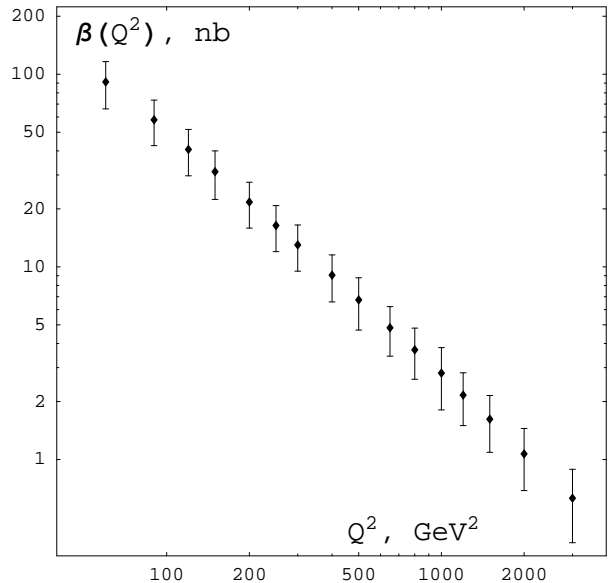}
\end{center}
\vskip -0.7cm
\caption{The hard pomeron residue $\beta(Q^2)$ at various $Q^2$.}
\label{bet}
\end{figure}

The $\gamma^*p$ total cross-section $\sigma^{\gamma^*p}_{tot}(W,Q^2)$ is related to the proton unpolarized structure function $F^p_2(x,Q^2)$ \cite{struc} through the 
well-known formula 
$$
\sigma^{\gamma^*p}_{tot}(W,Q^2)=\frac{4\pi^2\alpha_e}{Q^4}\frac{Q^2+4 m_p^2x^2}{1-x}F^p_2(x,Q^2)\,,
$$
where $x=\frac{Q^2}{W^2+Q^2-m_p^2}$, $\alpha_e$ is the electromagnetic coupling, $m_p$ is the proton mass, $W$ is the invariant mass of the produced hadronic state, and 
$Q^2$ is the incoming photon virtuality. Our reasoning is grounded on the phenomenological fact that in the lepton-proton deep inelastic scattering there is a wide 
kinematic range wherein $\sigma^{\gamma^*p}_{tot}(W,Q^2)$ can be satisfactorily described by means of a simple power expression: 
\begin{equation}
\sigma^{\gamma^*p}_{tot}(W,Q^2)\approx\beta(Q^2)\left(\frac{W^2+Q^2}{W_0^2}\right)^\delta,
\label{power}
\end{equation}
where $W_0\equiv 1$ GeV, $\beta(Q^2)$ is some unknown function of $Q^2$, and $\delta$ depends neither on $Q^2$ nor on $W$. The description details are presented in 
Figs. \ref{risstr} and \ref{bet} and Tab. \ref{dis}. The description quality is $\frac{\chi^2}{N_{DoF}}\approx 1.01$, $N_{DoF}=208$ (2 outlying points within the 
considered kinematic range were excluded from the fitting procedure, see Tab. \ref{out}). The produced estimation of $\delta$ is $\delta=0.317\pm 0.028$.

\newpage

\section*{The interpretation}

For exploration of the $W$-behavior of $\sigma^{\gamma^*p}_{tot}(W,Q^2)$ in the region $\{m_p^2,\,Q^2\}\ll W^2$ the Regge formalism \cite{collins} is useful (treating 
$F^p_2(x,Q^2)$ from the Regge phenomenology standpoint is not a new idea, {\it e.g.} see \cite{brandt}). In the framework of this formalism, we can represent 
the $\gamma^*p$ total cross-section (proportional, according to the optical theorem, to the imaginary part of the forward amplitude) in the following form: 
\begin{equation}
\sigma^{\gamma^*p}_{tot}(W,Q^2)=\frac{\Gamma_{\rm P}^{pp}(0)\Gamma_{\rm P}^{\gamma^*\gamma^*}(0,Q^2)}{W_0^2}\left(\frac{W^2+Q^2}{W_0^2}\right)^{\alpha_{\rm P}^h(0)-1}
+\; ... \;,
\label{regge}
\end{equation}
where $\alpha_{\rm P}^h(0)$ is the hard pomeron intercept, $\Gamma_{\rm P}(0)$ are the hard pomeron form-factors of the colliding particles at the zero of the transferred 
momentum, and ``...'' in the right-hand side corresponds to the combined contribution of Regge cuts and secondary poles. Within the explored kinematic range these terms 
are conjectured to be subleading corrections only. The theoretical arguments in favor of this conjecture are considered in detail in \cite{yndurain}, \cite{kaidalov}, and 
\cite{petrov}. Below we just restrict ourselves by some qualitative phenomenological reasoning.

The secondary reggeons make $\sigma^{\gamma^*p}_{tot}(W,Q^2)$ larger at lower energies and the absorptive corrections make $\sigma^{\gamma^*p}_{tot}(W,Q^2)$ smaller at 
higher energies. Consequently, the combined effect of these terms is in tempering the growth rate of $\sigma^{\gamma^*p}_{tot}(W,Q^2)$ with increasing $W$. The relative 
contribution of secondary poles decreases at higher $W$ (and, possibly, at higher $Q^2$). The relative contribution of Regge cuts decreases at higher $Q^2$, owing to 
the fact that the reggeon effective couplings to $\gamma^*$ should weaken at higher $Q^2$ due to asymptotic freedom. In such circumstances, at high enough values of the 
incoming photon virtuality and the $\gamma^*p$ collision energy these corrections could be neglected and, hence, the $W$-behavior of $\sigma^{\gamma^*p}_{tot}(W,Q^2)$ 
becomes universal. In other words, at higher $Q^2$ the growth rate of $\sigma^{\gamma^*p}_{tot}(W,Q^2)$ should tend from below to some constant value. This is the very 
pattern which takes place in experiment: the effective power $\lambda(Q^2)$, which characterizes the $W$-behavior of $\sigma^{\gamma^*p}_{tot}(W,Q^2)$ at high energies 
($\sigma^{\gamma^*p}_{tot}(W,Q^2)\sim W^{2\lambda(Q^2)}$), grows from $\sim 0.1$ at low $Q^2$ to $\sim 0.3$ at $Q^2>$ 60 GeV$^2$. Thus, the universal 
parameter $\delta$ in expression (\ref{power}) can be associated with the hard pomeron intercept: $\delta=\alpha_{\rm P}^h(0)-1=0.317\pm 0.028$. 

Note, that this purely phenomenological estimation is in agreement with the qualitative result obtained in the framework of the LO BFKL approximation, 
$\alpha^{\rm BFKL}_{\rm P}(0)-1\ge 0.3$ \cite{lipatov}, though the LO BFKL restriction is affected by the higher order corrections \cite{camici}. In other words, the 
value of $\alpha_{\rm P}^h(0)$ presented above is roughly consistent with the BFKL expectations, though these are not sufficiently accurate for a firm conclusion 
to be made.

\begin{table}[ht]
\begin{center}
\begin{tabular}{|l|l|l|l|}
\hline
Set of data       &  $\beta(Q^2)$, nb & Number of points &  $\chi^2$        \\
\hline
$Q^2=60$   GeV$^2$, 60  GeV $<W<$  250  GeV & 91.2 $\pm$ 25.1  &  7    &   3.3    \\
$Q^2=90$   GeV$^2$, 40  GeV $<W<$  250  GeV & 58.0 $\pm$ 15.4  & 10    &  12.0    \\
$Q^2=120$  GeV$^2$, 40  GeV $<W<$  250  GeV & 40.7 $\pm$ 11.0  & 11    &   1.2    \\
$Q^2=150$  GeV$^2$, 35  GeV $<W<$  250  GeV & 31.2 $\pm$  8.8  & 12    &  12.4    \\
$Q^2=200$  GeV$^2$, 35  GeV $<W<$  250  GeV & 21.7 $\pm$  5.8  & 16    &  10.4    \\
$Q^2=250$  GeV$^2$, 35  GeV $<W<$  250  GeV & 16.4 $\pm$  4.4  & 17    &  10.2    \\
$Q^2=300$  GeV$^2$, 30  GeV $<W<$  250  GeV & 13.0 $\pm$  3.5  & 21    &  24.1    \\
$Q^2=400$  GeV$^2$, $\sqrt{3Q^2}<W<$  250  GeV & 9.06 $\pm$ 2.48  & 18    & 27.7     \\
$Q^2=500$  GeV$^2$, $\sqrt{3Q^2}<W<$  250  GeV & 6.74 $\pm$ 2.04  & 17    & 15.7     \\
$Q^2=650$  GeV$^2$, $\sqrt{3Q^2}<W<$  250  GeV & 4.83 $\pm$ 1.39  & 18    & 20.6     \\
$Q^2=800$  GeV$^2$, $\sqrt{3Q^2}<W<$  250  GeV & 3.71 $\pm$ 1.10  & 15    & 12.2     \\
$Q^2=1000$ GeV$^2$, $\sqrt{3Q^2}<W<$  250  GeV & 2.81 $\pm$ 1.00  & 13    &  6.2     \\
$Q^2=1200$ GeV$^2$, $\sqrt{3Q^2}<W<$  250  GeV & 2.16 $\pm$ 0.66  & 15    & 20.8     \\
$Q^2=1500$ GeV$^2$, $\sqrt{3Q^2}<W<$  250  GeV & 1.62 $\pm$ 0.53  & 13    & 11.6     \\
$Q^2=2000$ GeV$^2$, $\sqrt{3Q^2}<W<$  250  GeV & 1.07 $\pm$ 0.38  & 11    &  7.9     \\
$Q^2=3000$ GeV$^2$, $\sqrt{3Q^2}<W<$  250  GeV & 0.63 $\pm$ 0.26  & 11    & 14.8     \\
\hline
\end{tabular}
\caption{The quality of description of the data on the $\gamma^*p$ total cross-sections by means of expression (\ref{power}) with $\delta=0.317$.}
\label{dis}
\end{center}
\end{table}
\begin{table}[ht]
\begin{center}
\begin{tabular}{|l|l|l|}
\hline
$Q^2$, GeV$^2$ & $W$, GeV & $\sigma^{\gamma^*p}_{tot}(W,Q^2)$, nb  \\
\hline
1000  &  67.5 & $33.7\pm 4.9$        \\
1000  &  81.8 & $62.7\pm 7.2$        \\
\hline
\end{tabular}
\caption{The outlying points excluded from the fitting procedure.}
\label{out}
\end{center}
\end{table}

The proposed estimation is significantly lower than the hard pomeron intercept values presented in the more complicated reggeon models \cite{donlan,desgrolard,martynov} 
with some secondary-reggeon contributions taken into account ($\alpha_{\rm P}^h(0)-1\simeq 0.4$) and higher than the corresponding values presented in \cite{lengyel}. 
However, these values were obtained through fitting to the old HERA data and they do not provide a satisfactory description of the higher-quality data \cite{struc}. 

\section*{Conclusions}

In conclusion, it would be appropriate to repeat the phenomenological arguments why the above-presented estimation of power $\delta$ in (\ref{power}), obtained via 
fitting to the data on the $\gamma^*p$ total cross-section in the region 60 GeV$^2<Q^2\ll W^2<60000$ GeV$^2$, should be associated with the hard pomeron intercept.
\begin{itemize}
\item The very fact that expression (\ref{power}) with universal $\delta$ provides a satisfactory description of the data within such a wide kinematic range is an 
undisguised hint about the adequacy of this simple approximation. 
\item At higher $Q^2$ the effective growth rate of $\sigma^{\gamma^*p}_{tot}(W,Q^2)$ tends to some constant value from below. Such an evolution is in accordance with 
the fact that in the presence of some hard scale the comparative contribution of absorptive corrections decreases due to the effect of asymptotic freedom.
\end{itemize}

As the hard pomeron is, presumably, the leading reggeon, with the highest Regge trajectory in the scattering region, so the hard-pomeron exchanges are expected to 
dominate in diffractive (small-angle) scattering of hadrons at ultra-high energies. Therefore, more or less precise estimation of the hard pomeron intercept is 
absolutely necessary for raising the predictive efficiency of the models describing the energy evolution of the corresponding cross-sections (for more detailed 
discussion, see, for example, \cite{donlan2} or \cite{donlan3}).

\end{document}